\documentclass[prl,twocolumn,aps]{revtex4}
\usepackage{graphicx}
\usepackage{natbib}
\usepackage{color}
\usepackage{amsmath}
\def\strutdepth{\dp\strutbox}
\def\nw#1{\strut\vadjust{\kern-\strutdepth\vtop to0pt{\vss\hbox to\hsize
{\hskip\hsize\hskip5pt$\leftarrow$\hss\strut}}}{\em #1}}

\begin{document}

\title{Surface Tension regularizes the Crack Singularity of Adhesive Contacts}

\author{Stefan Karpitschka$^{1}$, Leen van Wijngaarden$^{1}$, and Jacco H. Snoeijer$^{1,2}$}
\affiliation{
$^1$Physics of Fluids Group, Faculty of Science and Technology, Mesa+ Institute, University of Twente,
7500 AE Enschede, The Netherlands.\\
$^2$Mesoscopic Transport Phenomena, Eindhoven University of Technology, Den Dolech 2, 5612 AZ Eindhoven, The Netherlands}

\begin{abstract}
The elastic and adhesive properties of a solid surface can be quantified by indenting it with a rigid sphere. Indentation tests are classically described by the JKR-law when the solid is very stiff, while recent work highlights the importance of surface tension for exceedingly soft materials. Here we show that surface tension plays a crucial role even for stiff solids: it regularizes the crack-like singularity at the edge of the contact. We find that the edge region exhibits a universal, self-similar structure that emerges from the balance of surface tension and elasticity. The similarity theory provides a complete description for adhesive contacts, reconciling the global adhesion laws and local contact mechanics.
\end{abstract}

\date{\today}%

\maketitle

The adhesion between two solid bodies in contact is extremely common in nature and technology \cite{Luan2005,Tian2006,Christ2010,Zong2014}. Adhesive contacts are described by a classical law derived by Johnson, Kendall and Roberts (JKR) \cite{JKR71}, providing the benchmark to characterize elastic and adhesive material properties~\cite{Chaudhury1990,ChuPRL2005,Pussak2013}. Despite its success, JKR theory does not provide a complete description of the physics inside the contact. Namely, the elastic problem is considered without explicitly treating adhesive interactions, while furthermore the contact exhibits a crack-like singularity at edge. 
This issue was elegantly resolved in a macroscopic theory, where, in analogy to fracture mechanics, the elastic energy released by opening the crack is balanced by the work of adhesion \cite{Maugis1978,Greenwood1981,Barquins1988,Maugis1992,Chaudhury1996}.

\begin{figure}[t!]
	\centering\includegraphics[width=86mm]{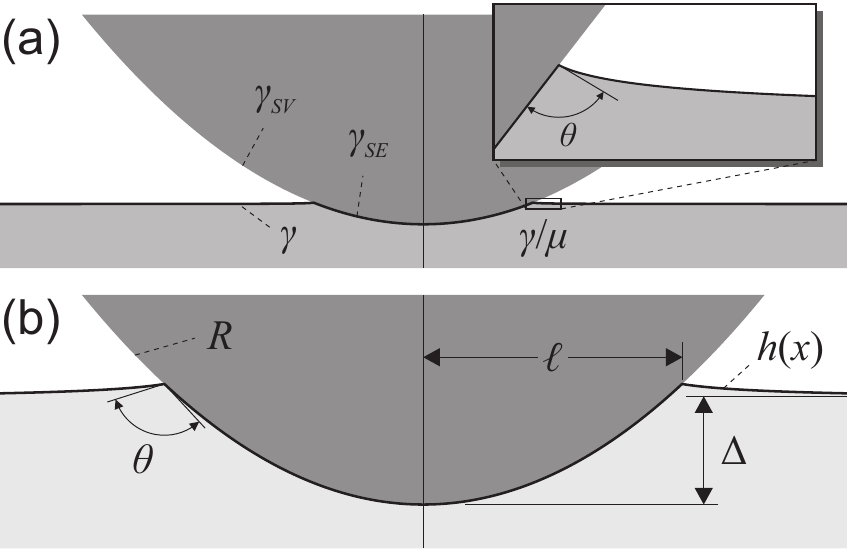}
	\vspace{-5 mm}
	\caption{\label{fig:sketch}Adhesive contact between an elastic layer and a rigid indenter. (a) Stiff contact, the elasto-capillary length $\gamma/\mu$ is much smaller than the contact size $\ell$. (b) Soft contact, $\gamma/\mu$ is comparable to $\ell$. Profiles $h(x)$ are solutions of our local contact theory (vertical scale stretched). The upper case is in the classical JKR regime, for which we identify a narrow, universal zone that is governed by a wetting angle $\theta$, regularizing the crack singularity (inset).}
	\vspace{-5 mm}
\end{figure}

Recent studies on very soft gels and rubbers provided a very new perspective on adhesive contacts~\cite{Long2012,StyleNatComm2013,SalezSM2013,Cao2014,HuiProc2015,LiuSM2015}. Just like fluids, these soft materials are highly susceptible to a Laplace pressure due to surface tension~\cite{MPFPP10,JXWD11,MDSA12,Limat12,ParetkarSM14}. This insight exposed a profound link between ``solid adhesion" and ``liquid wetting" \cite{Long2012,StyleNatComm2013,SalezSM2013,Cao2014,HuiProc2015,LiuSM2015}: both are adhesive contacts, but whether they are solid-like or liquid-like depends on the contact size ($\ell$ in Fig.~\ref{fig:sketch}) compared to the elastocapillary length $\gamma/\mu$, the ratio of surface tension $\gamma$ to shear modulus $\mu$. Liquid-like contacts were for example found for soft gels \cite{StyleNatComm2013} and nanoparticles \cite{Cao2014}.

Intriguingly, the limit of liquid wetting does \emph{not} suffer from the crack singularity: it is governed by a benign boundary condition, the wetting angle $\theta$. This stark contrast to the JKR-singularity, in combination with recent experiments \cite{Jensen2015}, once more provokes the question: What happens at the edge of adhesive contacts?

In this Letter we show that surface tension is crucial also for stiff, adhesive contacts. Our analysis reveals how surface tension regularizes the crack singularity on a scale $\gamma/\mu$, replacing it by a wetting boundary condition [cf. zoom in Fig.~\ref{fig:sketch}(a)]. Remarkably, we find that the contact exhibits at the edge a universal self-similar structure, described by a single similarity solution. The similarity theory offers a complete description of the physics inside the contact, including the crack, reconciling the global adhesion laws and the local contact mechanics.

\paragraph{Variational analysis: the wetting angle.---}The common treatment of adhesive contacts hinges on the energy released upon opening a crack \cite{Barquins1988,Chaudhury1996,HuiProc2015,LiuSM2015}. Here we take a different approach by posing the macroscopic free energy \footnote{We assume large deformations compared to the range of molecular interactions \cite{Maugis1992}, allowing macroscopic theory.} of the problem sketched in Fig.~\ref{fig:sketch}, and derive a boundary condition by direct variational analysis. This yields a wetting angle $\theta$, instead of the crack singularity.

As soft materials are nearly incompressible, we need to consider only the normal displacement $h(x)$ of the free surface of the elastic layer \cite{Musk,LiuSM2015}. The free energy of the problem then consists of an elastic functional, $\mathcal{F}_{el}[h]$, and capillary contributions due to the surface energies $\gamma$, $\gamma_{SV}$, and $\gamma_{SE}$ [cf. Fig.~\ref{fig:sketch}(a)]. The associated work of adhesion reads $W=\gamma+\gamma_{SV}-\gamma_{SE}$, representing the reduction in the surface energy when contact is created. The free energy needs to be minimized under the constraint that the elastic surface must comply with the shape of the indenter $f(x)$. This gives the constraint $h(x) = f(x) + \Delta$ over the range $-\ell \leq x \leq \ell $, where $-\Delta$ is the distance by which the solid is indented. Taking also into account the work done by the external (two-dimensional) load $f_{2D}$, the functional to be minimized is (using the symmetry $x \rightarrow -x$),
\begin{widetext}
\begin{eqnarray}\label{eq:F}
\mathcal{F}[h(x);\ell,\Delta]&=&\mathcal{F}_{el}[h(x)] + \int_\ell^\infty dx \, \gamma \left(1+h'^2 \right)^{1/2} + \int_\ell^\infty dx \, \gamma_{SV} \left(1+f'^2 \right)^{1/2} + \int_0^\ell dx \, \gamma_{SE} \left(1+f'^2 \right)^{1/2} \nonumber \\
&&+ \int_0^\ell dx \, p(x) \left\{h(x) - \left[f(x)+\Delta \right] \right\} + \lambda \left\{ h_+(\ell) - \left[f(\ell)+\Delta \right]   \right\} + \frac{1}{2} f_{2D} \Delta.
\end{eqnarray}
\end{widetext}
The integrals in the upper line give the surface areas multiplied by their interfacial energies. Compliance in the contact zone is imposed by the continuous Lagrange multiplier $p(x)$. Without imposing the slope outside the contact, we require $h(x)$ to be continuous at $x=\ell$, which is why we introduce the Lagrange multiplier $\lambda$ to ensure compliance just outside the contact [$h_+(\ell) = \lim_{x\to\ell^+} h(x)$]. The resulting variational problem resembles that of a liquid drop on an elastic layer~\cite{LUUKJFM14}.

The degrees of freedom of the problem thus are $h(x),  \ell, \Delta$, which should minimize the functional (\ref{eq:F}). As shown in the Supplementary Information~\cite{supplementary}, the variations define the elastic problem as [eliminating $p(x),\lambda$]:

\begin{eqnarray}
h(x)&=&f(x)+\Delta , \quad  \quad |x| < \ell, \\
\sigma(x) &=& \frac{\gamma h''}{\left(1+h'^2 \right)^{3/2}}, \quad |x| > \ell, \label{eq:laplace} \\
\gamma \cos \theta &=& \gamma_{SV} - \gamma_{SE}, \quad x= \pm \ell, \label{eq:young}
\end{eqnarray}
relating to the total force as

\begin{equation}
f_{2D} =   \frac{2\gamma h'_+}{\left(1+h_+'^2 \right)^{1/2}} - \int_{-\ell}^\ell dx \, \sigma(x).
\end{equation}
Here $\sigma \equiv \delta \mathcal{F}_{el}/\delta h$ is the elastic normal stress, which supports the capillary pressure outside the contact (\ref{eq:laplace}).

The crux of the analysis is the appearance of Young's wetting angle $\theta$, defined in Fig.~\ref{fig:sketch}: it serves as the boundary condition emerging from the variation of $\ell$. Like in elastic wetting \cite{JXWD11,Limat12,LUUKJFM14}, surface tension thus dominates over elasticity at the contact line and replaces the crack by a ``wedge" geometry. Such a wetting-like boundary condition is also hypothesized by Jensen \emph{et al.}~ \cite{Jensen2015}, who experimentally found that a gel layer approaches the indenter at a well-defined angle that is independent of the global geometry. Importantly, from a theoretical point of view, the adhesive properties are now contained in the theory, since (\ref{eq:young}) can be expressed as $W=\gamma(1+\cos \theta)$.

\paragraph{Cylindrical Indenter.---} Within linear elasticity, the stress and displacement of the free surface of an incompressible thick layer are related by the integral~\cite{Musk},

\begin{equation}\label{eq:dim}
\sigma(x) = - \frac{2\mu}{\pi } \int_{-\infty}^\infty dt \, \frac{h'(t)}{t-x},
\end{equation}
valid for all $x$. We now focus on a cylindrical indenter of radius $R$, approximated by $f(x) = \frac{x^2}{2R}$. Strict validity of linear elasticity requires small strains, i.e. $h'^2 \ll 1$, and thus requires that $\theta$ be close to $\pi$. This simplifies the boundary condition to $h'_- - h'_+ = (2W/\gamma)^{1/2}$, showing that $W$ enters in the form a slope discontinuity. It is advantageous to introduce dimensionless variables:

\begin{equation}
X = x/\ell, \quad H = h R/\ell^2, \quad F_{\rm 2D} = \frac{f_{2D}R}{\pi \mu \ell^2}, 
\end{equation}
where the external load was scaled by the result for nonadhesive (Hertz) contacts. It turns out convenient to define the following dimensionless numbers: 

\begin{equation}\label{eq:AS}
S =  \frac{\pi \gamma}{2\mu \ell}, \quad A=\left(\frac{2WR^2}{\gamma \ell^2 }\right)^{1/2}.
\end{equation}
$S$ represents the influence of surface tension, comparing the elasto-capillary length $\gamma/\mu$ to the width $\ell$ of the contact. $A$ represents the work of adhesion, dictating the wetting angle. 
Following \cite{LiuSM2015}, we note that outside the contact $\sigma(x)=\gamma h''(x)$, and one can reduce (\ref{eq:dim}) to \cite{supplementary}:

\begin{equation}\label{eq:eq2}
S H''(X) = - \int_1^\infty dT \, \frac{2T \,H'(T)}{T^2-X^2} - \left[ 2 + X \ln |\frac{X-1}{X+1}| \right], 
\end{equation}
valid for $X>1$, outside the contact. 

The integral equation (\ref{eq:eq2}) completely defines the adhesive contact problem. Surface tension $S$ is represented explicitly, while the adhesion parameter $A$ appears as the slope discontinuity boundary condition, $H'(1)=1-A$. The examples in Fig.~\ref{fig:sketch} are actual numerical solutions with $S\neq 0$. Importantly, the boundary condition \emph{cannot} be imposed without surface tension, $S=0$, for which the solution exhibits the crack-singularity around $X=1$,

\begin{equation}\label{eq:JKRexpand}
H_0'(X) \simeq X -K (X-1)^{-1/2}.
\end{equation}
Here $K$ is essentially the (scaled) stress intensity factor, which for the cylinder reads $K=(1-F_{2D})/2^{3/2}$ \cite{Musk}.  

\begin{figure}[t!]
	\centering\includegraphics[width=86mm]{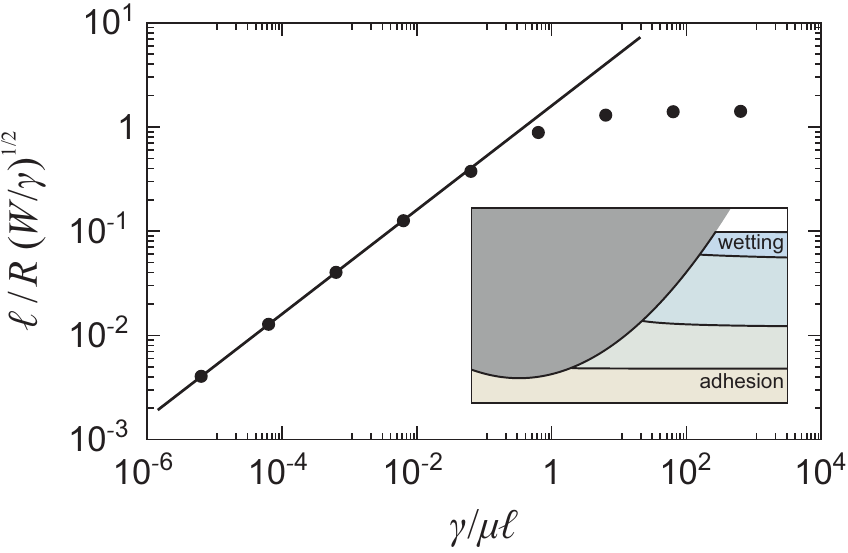}
	\vspace{-5 mm}
	\caption{\label{fig:indent} Numerical solutions to the local contact mechanics equation (\ref{eq:eq2}) without external load ($f_{2D}=0$). The contact width $\ell$ (made dimensionless using the scaling of $A$) is shown as a function of the surface tension $\gamma$ (made dimensionless using the scaling of $S$). For small $\gamma/\mu\ell$, we recover the JKR scaling [solid line, (\ref{eq:JKRscaling})], while for large $\gamma/\mu \ell$ the width saturates. The inset shows the corresponding profiles.}
	\vspace{-5 mm}
\end{figure}

\paragraph{Results for vanishing load.---} To illustrate that our local theory indeed provides a complete description, we present numerical solutions to (\ref{eq:eq2}) with boundary condition $H'(1)=1-A$, for the case of vanishing load. Figure~\ref{fig:indent} shows how the width $\ell$ varies with surface tension $\gamma$ [both scaled according to $A$ and $S$, cf. (\ref{eq:AS})]. For small $S$ we perfectly recover the result for ``solid adhesion" (solid line) \cite{Barquins1988,Chaudhury1996}, which in the present notation reads 

\begin{equation}\label{eq:JKRscaling}
\ell =  \left( \frac{8R^2W}{\pi \mu} \right)^{1/3} \quad \Rightarrow \quad AS^{1/2} = 2^{-3/2} \pi. 
\end{equation}
By contrast, for large surface tension (or equivalently vanishing $\mu$), the contact width becomes independent of $S$ and the curve saturates to the value expected for ``liquid wetting". This is in agreement with the analysis based on energy release rate by Liu \emph{et al.}~\cite{LiuSM2015}, but here derived from an analysis of the entire contact -- including the crack region and the angle $\theta$. The result in Fig.~\ref{fig:indent} also concurs with experiments \cite{StyleNatComm2013} and simulations \cite{Cao2014}. 

\begin{figure}[t!]
	\centering\includegraphics[width=86mm]{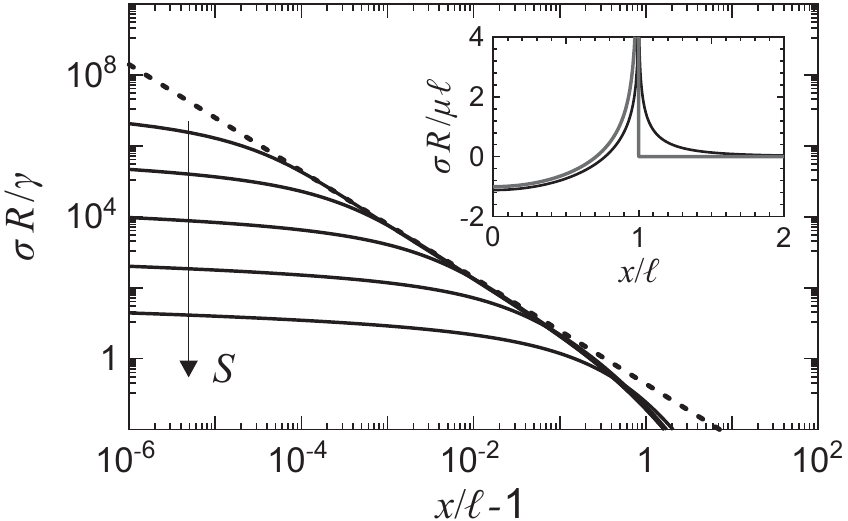}
	\vspace{-5 mm}
	\caption{\label{fig:stress} Stress profiles $\sigma$ outside the contact with $f_{2D}=0$, for $S=10^{-4}, 10^{-3}, 10^{-2}, 10^{-1},1$. The stress exhibits a -3/2 scaling (dashed line), regularized by a logarithmic zone as $x \rightarrow \ell$. Inset: Stress profile on the scale of the contact. Grey line is JKR profile ($S=0$), black line a typical profile with surface tension ($S=1$).}
	\vspace{-5 mm}
\end{figure}

The present theory has the additional merit that it reveals the stresses near the contact edge. The inset of Fig.~\ref{fig:stress} shows a stress profile on the scale of the contact ($S=1$, black line). The stress extends both inside and outside the contact and exhibits a weak, logarithmic singularity associated to the slope discontinuity~\footnote{The logarithmic stress divergence due to a small discontinuity in $h'$ lies within the realm of linear elasticity \cite{Johnson}. The scaling (\ref{eq:stresssim}) confirms that $\sigma \sim \mu$ only within an asymptotically small region near the contact, $X-1 \lesssim S \exp(-\sqrt{\gamma /W})$, where we note that the present analysis assumes $(W/\gamma)^{1/2} \ll 1$.}. This is very different from the JKR theory ($S=0$), which has the square root singularity at the inside and vanishing stress outside the contact (inset, grey line). The main panel in Fig.~\ref{fig:stress} shows further details of the stress when approaching $x\rightarrow \ell$ from the outside of the contact. The various curves correspond to different $S$. One observes the appearance of a -3/2 power law, which is logarithmically smoothened at small distances where surface tension becomes dominant. The various profiles all look very similar. The cross-over observed in Fig.~\ref{fig:indent} can thus be explained from the physics inside the contact: the transition from ``adhesion" to ``wetting" appears when the influence of surface tension, the smoothing zone, becomes comparable to the size of the contact, i.e. $\gamma/\mu \sim \ell$. 

\paragraph{Universality of the edge region: similarity analysis.---} We now expose the self-similar nature of the edge region. Since the extent of surface tension is bounded to a thin region that scales with $S$, we introduce a boundary layer to regularize the singular crack solution~\footnote{A similar approach was used for the elasto-hydrodynamic lubrication of Hertzian contacts in~\cite{SnoeijerPOF2013}}. We therefore propose the similarity form, valid for $S\ll 1$,

\begin{equation}\label{eq:simform2}
H(X) = AS \mathcal{H}(\zeta)  + \frac{1}{2}X^2, \quad \zeta = \frac{X-1}{S}.
\end{equation}
Here $\mathcal{H}(\zeta)$ is a universal function describing the inner region of the crack. As detailed in the Supplementary \cite{supplementary}, inserting (\ref{eq:simform2}) in (\ref{eq:eq2}) gives a closed equation for $\mathcal{H}(\zeta)$:

\begin{equation}\label{eq:magic}
\mathcal{H}''(\zeta) = - \int_0^\infty d\zeta' \, \frac{\mathcal{H}'(\zeta')}{\zeta'-\zeta},
\end{equation}
complemented by the boundary condition $\mathcal{H}'(0)=-1$.  

\begin{figure}[t!]
	\centering\includegraphics[width=86mm]{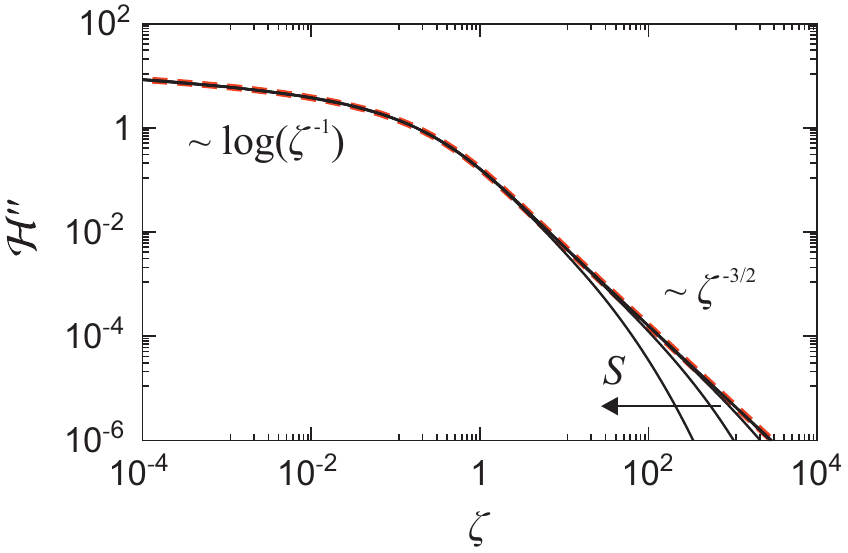}
	\vspace{-5 mm}
	\caption{\label{fig:similarity} The similarity function $\mathcal{H}''(\zeta)$ describing the crack region (red, dashed). Superimposed are the stress profiles of Fig.~\ref{fig:stress}, scaled according to (\ref{eq:stresssim}), confirming the self-similarity.}
	\vspace{-5 mm}
\end{figure}

The numerical solution to (\ref{eq:magic}) is represented in Fig.~\ref{fig:similarity}, showing $\mathcal{H}''(\zeta)$ as the red dashed line. Owing to the capillary stress relation, $\sigma=\gamma h''= \gamma A \mathcal{H}''/RS$, the similarity Ansatz in fact predicts a collapse of all stress profiles:

\begin{equation}\label{eq:stresssim}
\sigma(x) = 
%\frac{\gamma A}{RS} \mathcal{H}''(\zeta) 
\frac{2^{3/2}\mu}{\pi} \left(\frac{W}{\gamma}\right)^{1/2} \mathcal{H}''(\zeta).
\end{equation}
Figure~\ref{fig:similarity} shows that the scaled stress profiles of Fig.~\ref{fig:stress} perfectly collapse on the predicted dashed line, illustrating the validity of the similarity theory.

\begin{figure}[t!]
	\centering\includegraphics[width=86mm]{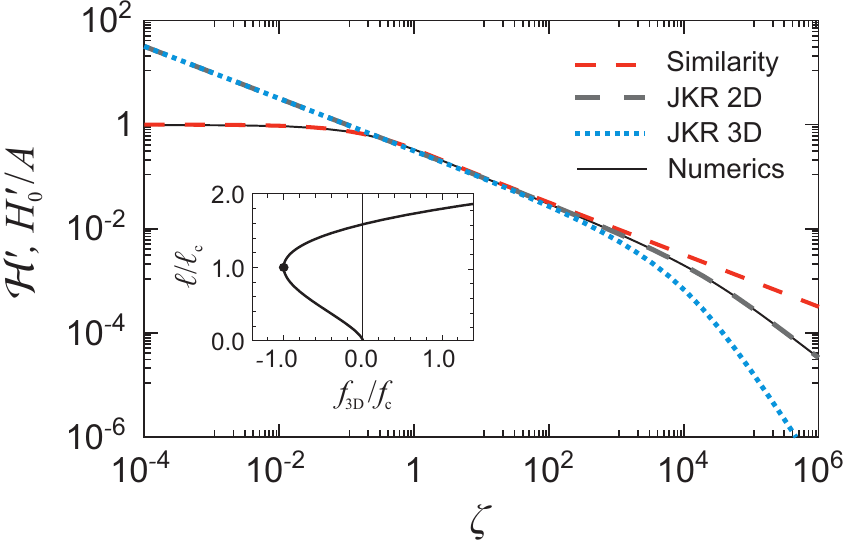}
	\vspace{-5 mm}
	\caption{\label{fig:JKR} Matching the asymptotes of the similarity solution $\mathcal{H}'$ (red, dashed), to the singular outer solutions $H'_0/A$ for the cylinder (gray dashed) or sphere (blue dotted). The solid black line shows the 2D numerics: it follows $H_0'$ on large scales, switching to $\mathcal{H}'$ on small scales. The matching was done for $S=10^{-4}$, while the external load was set to the critical value for pull-off (inset, circle). Inset: $\ell$ versus $f_{3D}$ given by Eq.~(\ref{eq:match3D}), scaled by the critical values $f_c,\ell_c$.}
	\vspace{-5 mm}
\end{figure}

\paragraph{Adhesion laws: matched asymptotics.---}
To complete the analysis, we demonstrate how the adhesion laws are recovered. We observe that $\mathcal{H}'\simeq -1/(\pi \zeta^{1/2})$ for large $\zeta$, and this behavior can be matched to the square root divergence of (\ref{eq:JKRexpand}). This is illustrated in Fig.~\ref{fig:JKR}: when the effect of surface tension is small, $S\ll 1$, one observes an overlap between the inner solution ($\mathcal{H}'$, red dashed) and the outer solution ($H_0'$, gray dashed). Equating the prefactors of the asymptotes gives,

\begin{equation}\label{eq:match2D}
K=\pi^{-1} AS^{1/2} \quad \Rightarrow \quad 1-\frac{f_{2D}R}{\pi \mu \ell^2} = \left( \frac{8WR^2}{\pi \mu \ell^3}\right)^{1/2},
\end{equation}
where in the second step we used the dimensionless stress intensity factor $K=(1-F_{2D})/2^{3/2}$. 

The matching condition (\ref{eq:match2D}) coincides with the adhesion law for the cylindrical indenter, previously derived from a balance of the energy release rate and $W$ \cite{Barquins1988,Chaudhury1996}. For the case $f_{2D}=0$ it reduces to (\ref{eq:JKRscaling}), while without adhesion it gives the Hertz law between the size of the contact $\ell$ and the applied load. The similarity theory gives a physical description of this result: the adhesion imposes a wetting angle at the edge of the contact, which by means of a boundary layer is connected to the elastic displacements on the much larger scale of the contact.

Analogously, the same matching gives the JKR-law for the \emph{spherical} indenter. The boundary layer is asymptotically thin compared to the contact radius, so the physics at the edge is quasi one-dimensional (see also \cite{SnoeijerPOF2013}). The spherical contact is therefore governed by the same universal form $\mathcal{H}(\zeta)$ and again requires a matching to the singular $H_0$ (Fig.~\ref{fig:JKR}, blue dotted). Evaluating $K$ for the spherical indenter indeed gives the JKR-law~\cite{JKR71}:
%$K=2^{2/3}(1-F_{3D})/(3\pi)$, where $F_{3D}$ is the external force $f_{3D}$ scaled by the Hertz load, to recover the classical JKR-law~\cite{JKR71}:

\begin{equation}\label{eq:match3D}
K=\pi^{-1} AS^{1/2} \quad \Rightarrow \quad 1-\frac{3 f_{3D}R}{16 \mu \ell^3} = \left( \frac{9\pi WR^2}{8 \mu \ell^3}\right)^{1/2}.
\end{equation}
For completeness, we illustrate this famous relation between $\ell$ and the load $f_{3D}$ in the inset of Fig.~\ref{fig:JKR}. The outer solutions shown in the main panel of Fig.~\ref{fig:JKR} were selected to correspond to the case where the external force reaches its minimum (negative) value, which is the critical point where the indenter is pulled off from the elastic layer.

\paragraph{Outlook.---}

The present analysis provides a complete unification of solid adhesion and wetting: both are governed by a wetting angle, and this regularizes the crack singularity. While the elastic solution given here is restricted  to linear elasticity, $(W/\gamma)^{1/2} \ll 1$, the boundary condition (\ref{eq:young}) equally applies nonlinearly -- as also suggested by recent experiments \cite{Jensen2015}. A noteworthy nonlinear effect at the contact line is that the solvent can be separated from the gel in the vicinity of the edge \cite{Jensen2015}, in particular close to complete wetting \cite{Mehrabian2015}. In a broader context, we hypothesize that self-similarity could be a generic feature of fracture in the presence of surface tension \cite{Wang2008,LiuSM2014}, as is the case for fracture of viscous liquids \cite{E01,LRQ03}.

\acknowledgments{We thank B. Andreotti, C.-Y. Hui, E. Raphael and T. Salez and  for discussions. SK acknowledges financial support from NWO through VIDI Grant No. 11304. JS acknowledges financial support from ERC (the European Research Council) Consolidator Grant No. 616918.}

%\bibliography{Jacco_elastocap}
%%\bibliographystyle{jfm} %the RSC's .bst file
%%\bibliographystyle{rsc} %the RSC's .bst file
%%}
%\iffalse
%\fi

\end{document}